# Using Images to create a Hierarchical Grid Spatial Index

Lukasz A. Machowski, and Tshilidzi Marwala, *Member, IEEE*

*Abstract*— This paper presents a hybrid approach to spatial indexing of two dimensional (2D) data. It sheds new light on the age old problem by thinking of the traditional algorithms as working with images. Inspiration is drawn from an analogous situation that is found in machine and human vision. Image processing techniques are used to assist in the spatial indexing of the data. A fixed grid approach is used and bins with too many records are sub-divided hierarchically. Search queries are pre-computed for bins that do not contain any data records. This has the effect of dividing the search space up into non rectangular regions which are based on the spatial properties of the data. The bucketing quad tree can be considered as an image with a resolution of 2x2 for each layer. The results show that this method performs better than the quad tree if there are more divisions per layer. This confirms our suspicions that the algorithm works better if it gets to "look" at the data with higher resolution images. An elegant class structure is developed where the implementation of concrete spatial indexes for a particular data type merely relies on rendering the data onto an image.

## I. Introduction

THIS paper sheds new light on the way in which spatial indexing is considered and performed. It begins by giving a brief overview of two common spatial indexing techniques. It then gives evidence to suggest that an analogous situation exists in the human vision system and therefore opens the door to using image processing techniques to deal with the spatial indexing problem. The design and implementation of this spatial index is discussed and the results are presented and analyzed. A brief description of the further work explains where this research is headed.

## II. Literature Review

### A. Spatial Indexing

With the advent of high performance computing, there are an increasing amount of datasets that contain a significant spatial component to them. Being able to perform spatial queries on this data is only feasible if there is a way to manage the large quantities of data. Database indexing is a technique used to speed up searches for data by creating a searchable catalogue of the data based on a unique key. Spatial indexing uses the spatial coordinates of the data to create the searchable catalogue. This has the effect of prioritizing searches through the data, based on the spatial extent from the query point. Since one of the most recent developments in database technology is the addition of spatial data types, the spatial indexing methods that databases rely on are becoming increasingly more important [1].

The most typical spatial indexing scheme makes use of a divide-and-conquer approach where the original domain space is broken down into several regions. These regions are in turn divided up as necessary to form a hierarchical tree that can be traversed when searching for records. One of the most common examples of this spatial indexing technique is the *quad tree* (and region quad tree) for two dimensions and the *oct tree* for three dimensions [1]-[3] which create a recursive decomposition of space [2]. An alternative to the hierarchical data structure is the fixed grid or cell method, which is popular amongst cartographers. The advantage of this method is that it is easy to do lookups and adding or deleting records from the data structure is simple. The disadvantage of this method is that it is only suited to uniformly distributed data which is not typically the case when dealing with geographic information such as road network data for an entire country. The difficulty with hierarchical spatial indexes is in partitioning and grouping the records [1]. This can be done either as a batch process once all the records are available, or it can be done as the records are being added. Removing records from certain regions may get complicated and this limits the usefulness of the structure to dynamic information. There is also an overhead in memory and access time which must be taken into account when using hierarchical spatial indexes. An excellent review of the various spatial indexing techniques as well an extensive description and analysis are found in [2].

### B. Spatial Indexing Methods

In order to simplify the problem domain for this paper, we only consider the spatial indexing of points in two-dimensional space. This delimitation is acceptable since most other geometric primitives can be expressed as points and typical real world datasets are generated from geospatial data which is approximately 2D. We also only deal with the updating of the spatial indexes as a batch process in order to illustrate the key concepts presented in this paper.

*1) Fixed Grid Method*

The fixed grid method is one of the simplest spatial indexing methods to implement. It simply divides up the extents of the data into regular sized grid bins. Each bin maintains a list of all the data records that fall within the bin.

Manuscript received March 1, 2006. This work was supported in part by Storm Logistics, South Africa (www.profilerxp.com).

L. A. Machowski is with Storm Logistics and currently doing his PhD at the School of Electrical and Information Engineering, University of the Witwatersrand, Republic of South Africa (e-mail: lukem@profilerxp.com; l.machowski@ee.wits.ac.za).

T. Marwala, is with the School of Electrical and Information Engineering, University of the Witwatersrand, Republic of South Africa (e-mail: t.marwala@ee.wits.ac.za).

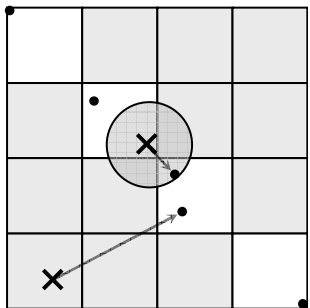

Fig. 1. Search required near the edge of a bin. Some bins are empty.

Searching for the nearest record involves finding the bin at the query location and iterating through each record in the list to find the nearest match. Resolving the bin indices is an *O(1)* operation because all we have to do is divide the offset of the query point from the corner of the spatial index extents, by the width of each bin.

For a uniformly distributed set of points, we expect the average search cost to be:

$$\bar{c} = \frac{n}{(x \times y)} \quad (1)$$

Where n is the total number of records, x is the number of divisions in the X dimension and y is the number of divisions in the Y dimension. Unfortunately, this equation is only valid near the centers of the bins because of the effect that occurs at the edges as illustrated in Fig 1.

For points that are near the edge of a bin, one must consider the points that are in the adjacent bins in order to ensure that the search result is indeed the closest record to the query point. The negative effect of this is compounded once we attempt to index data that is highly non-uniform. If the data has regions where there are gaps, there is a good chance that several of the bins will be empty. Querying for the nearest record at these locations will incorrectly return no results. It is therefore necessary to apply some heuristic to search through adjacent bins that are near the query point. This paper presents a neat solution to this problem and is discussed later.

*2) Quad Tree*

As discussed in [2], there are a number of different spatial indexing methods that can be classified as quad trees. The overall idea is that the region of interest is divided up into quadrants which are further sub-divided recursively until a set number of records are present in each quadrant. This is a hierarchical technique that can be thought of as processing at multiple resolutions, because the spatial extents of the quadrants are continually decreasing as the search depth in the tree increases. This paper mainly deals bucketing methods where the data records are added to buckets (or bins) that are defined by the extents of the quad tree quadrants [1]. An example quad tree is shown in Fig 2.

The problem with the quad tree is that one needs to travel several layers deep before one reaches a high enough resolution that is suitable for indexing through large amounts of data. Also, it becomes slightly more complicated to perform range searches because of the tree structure that needs to be taken into account when looking at adjacent bins.

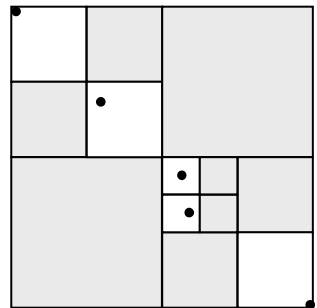

Fig. 2. An example of a bucketing quad tree for a small set of points. The maximum number of records in a bucket is 1.

### III. BACKGROUND

The inspiration for this spatial indexing technique comes from two sources. The first is that humans are very adept at visually searching through large sets of data while being able to filter out irrelevant details. The second is that most of the existing spatial indexing techniques can astonishingly be thought of as using images to perform the spatial searches visually. The techniques might not all use regular grid-like images, but their effect can definitely be considered as an image processing operation. A little more clarification is in order.

*A. Humans as excellent spatial indexers*

The most widely accepted theory of spatial vision is that of the multichannel model [4] developed by Enroth-Cugell and Robson [5] and Campbell [6]. This theory proposes that the visual system processes the retinal image simultaneously at several different spatial scales [4]. This is consistent with the type of data that needs to be processed in the real world, which is often made up of different levels of detail. The physiological evidence for this ability is in the size of a neuron's receptive field in each stage of early vision [4]. The neurons that process the raw signals from the photoreceptor cells have varying receptive field sizes and therefore we are able to detect a wide range of detail. It is believed that these various-scale outputs along with information from other channels are combined by the higher vision processes into our interpretation of a scene.

The impulse response of individual neurons falls-off as the signal moves further away from the centre of the neurons receptive field. This characteristic is highly desirable and essential for performing spatial queries on sets of data.

*B. Traditional Techniques as Image Operations*

One of the most common 2D spatial indexing techniques is the Quad-Tree as discussed earlier. If one considers that the quad tree is a pyramid of images, then one realizes that each layer is only represented by a 2x2 image. The same concept can be applied to other common spatial indexing techniques and one soon realizes that decisions on what search path to follow are based on a very limited view of the data. Imagine if all that we could see was 4 pixels at a time!

In this paper, we propose that one considers the problem of spatial indexing as an image processing operation which gets performed on multi-resolution representations of the underlying data. The process of inserting data into the spatial index is thought of as rendering the data onto an image, which is the regular grid at each hierarchical level. This means that we can use the wealth of knowledge and the abundance of algorithms that are available for drawing geometric primitives in order to insert the data into the spatial index. One can also use the concept of alpha-blending as a way of creating a histogram of regions that contain a large number of records. If we render the data with an additive drawing mode, then regions with a bright (high) color are known to contain large amounts of data records.

Considering the spatial index as a set of images also helps us when performing range and nearest point queries. This is because we are able to use the vast amount of morphological operators and other techniques that are available for image processing [7]-[10]. One of the fundamental concepts of morphological operators is the idea of neighborhoods and connectivity. Given a 5x5 image shown in Fig. 3, we can describe the 1-neighbourhood of a pixel as the set of pixels that are touching the centre pixel. In our definition, we consider the diagonal pixels as being connected as well. The $n$-neighborhood is therefore, the set of pixels that are $n$ pixels away from the centre pixel. The morphological operators make use of the neighboring pixels to decide on the value for the centre pixel. In fact, many 2D image filters are defined as kernels which are convolved with an image to perform complex operations such as blurring, sharpening, opening and closing [7][8].

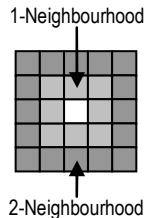

Fig. 3. The 1-(light) and the 2-(dark) neighborhood of a pixel.

Anderson and McCartney have shown that using images (or diagrams) can be very effective for performing several complex spatial database queries [10]. They use logical operators on 2D diagrams to perform the search queries. This paper extends their idea by using a hierarchical set of images to perform the spatial indexing of the data.

IV. METHOD

The spatial index described in this paper is designed as a set of Object-Oriented classes in C# for Microsoft .NET V2. The design makes use of inheritance, polymorphism and interfaces to achieve an elegant and extensible solution to the problem. The use of generics is not necessary to implement the spatial index.

The spatial index is implemented as a class hierarchy as shown in Fig. 4.

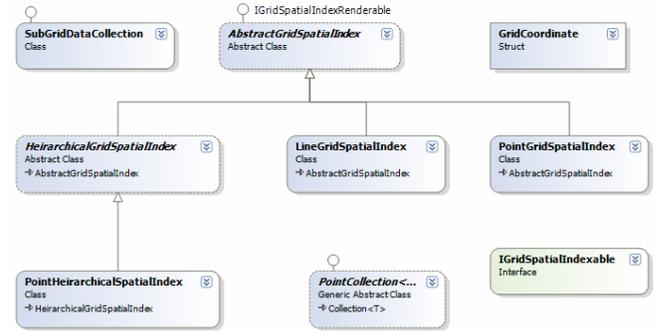

Fig. 4. Class Hierarchy for the Spatial Index

The *AbstractGridSpatialIndex* is the base class for all grid spatial indexes. This class contains all the common functionality for indexing data in a grid. It also performs the bulk of the indexing and spatial queries. This design allows us to have a solid and consistent implementation for the grid spatial index while allowing a variety of sub classes to implement different indexing behaviors. The class is declared as abstract so that sub classes are forced to implement the abstract methods. In order for the spatial index to be useful for indexing many types of data, it is necessary to make the abstract class accept a very general data type. For this reason, an interface is used instead of a concrete class and it is described in the next section.

*A. IGridSpatiaIndexable*

This is the interface that needs to be implemented by data collections in order for them to be indexed in a grid spatial index. The reason for using an interface is so that the spatial index does not limit the class structure that may be indexed. It allows arbitrary class hierarchies to exist for the data collections as long as the class implements the methods required for performing spatial indexing. The interface declaration also defines the minimal functionality required for indexing data in a grid spatial index. This interface is shown in Fig. 5.

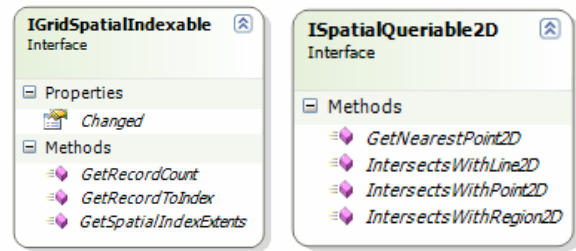

Fig. 5. *IGridSpatialIndexable* is the minimal interface required for spatial indexing in a grid. *ISpatialQueriable2D* is the minimal interface for data that can be indexed.

Note that the interface includes properties. This is a feature of the .NET framework and it allows interfaces to declare field-like elements that are implemented with getter and setter methods. If your language does not support this feature then the interface would merely have the corresponding getter and setter methods to replace the properties.

The spatial index has been designed to index only integer values. This scheme provides a good trade-off between generality for multiple applications and it also allows complex data access schemes to be spatially indexed. It is therefore the role of the *IGridSpatialIndexable* object to supply the mapping between an index and the actual record to be indexed. Concrete spatial index classes must implement the *GetRecordToIndex()* method of the interface. This method gets passed the index of the record to process and a temporary object of the type being indexed. The method must get the information for that specific record and return it to the spatial index. The temporary object that is passed to the method allows one to perform arbitrary calculations on the data (such as coordinate transformations) without having to create hundreds of transient objects for this process. This object reuse improves performance considerably, especially for cases where the record itself has to be converted into a form that can be indexed. The record to be indexed also has to implement the *ISpatialQueriable2D* interface, which defines the methods shown in Fig. 5.

It is necessary for the spatial index to get the extents of the data being indexed (*GetSpatialIndexExtents()*). This is so that the initial grid spatial index can be generated. A method that returns the total number of records (*GetRecordCount()*) is required for the spatial index to know how many records to index. The interface also has a Boolean property (*Changed*) which flags whether the data has changed. The spatial index uses this flag to rebuild itself whenever a spatial query is about to be run. With the *IGridSpatialIndexable* interface, we are able to represent a collection of data that can be indexed hierarchically or in only one layer of a grid index.

### B. AbstractGridSpatialIndex

This class takes *the IGridSpatialIndexable* collection and the number of divisions for the initial spatial grid as parameters to its constructor. It contains two grids of integer lists. The grid is implemented as a 2D array of integer lists. The first grid holds in each bin, the record indices that fall inside that bin. This grid is the result of rendering all the data to an image and saving which records were rendered to the pixels. Any bins that do not have data are set to null. Every integer list that is unique for this grid is maintained in a dictionary where the integer list is the key and the corresponding grid coordinate is the value. This allows an efficient lookup of the grid coordinates for a particular integer list. The second grid holds a duplicate of the rendered list but all the null bins are set to point to the integer list that contains the nearest record to the centre of the bin. This is a method of pre-computing approximate results to the problem discussed above for empty bins in the fixed grid spatial index. Rebuilding the spatial index is done when the data is flagged as being changed and may be described as the following high level process:

1. Get the spatial data-extents.
2. Get the number of records to index.
3. Clear the unique-list dictionary.
4. Create the bins for the lists.
5. Calculate the bin sizes.
6. Allow descendant classes to perform extra processing before the index is rebuilt.
7. Render the records into the grid (nulls where there is no data).
8. Create a shallow copy of the rendered lists.
9. Fill in the gaps by finding the integer lists with the nearest record to the centre of the bin.
10. Allow descendant classes to perform extra processing after the index is rebuilt.
11. Flag that the data has been processed and watch for further changes.

The class also has several protected helper methods to assist descendant classes to render their data correctly to the bins. These are in the form of efficient point, line and area rendering methods that add the indices of the records into the integer lists in the grid. This design means that concrete descendant classes only need to implement two methods for the spatial index to work, namely *RenderRecordsToLists()* and *CreateRecordInstance()* (which makes the temporary record described previously). It is evident from these two methods that we have successfully managed to abstract out all the spatial indexing functionality from the data rendering functionality. This means that the developer of concrete sub classes only has to program how to render the data to a grid (which is essentially the same as rendering the data to an image or the screen).

### C. Searching

The efficiency of a spatial index lies in its role as a pruning device for searching that is done [2]. In order to solve the empty-bin problem discussed earlier, we propose a solution that pre-computes the bin with the nearest record to the centre of each of the empty bins (step 9 above). This has the effect of creating non-rectangular regions that all point to the same integer list. This is a very desirable effect because the grid is partitioned into arbitrary regions that depend entirely on the data. Most other algorithms partition the search space into strict rectangular segments which are not well suited to real-world data. Fig. 6 shows an example of this partitioning.

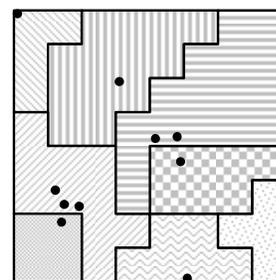

Fig. 6. The partitioning of reused bins after gaps in the grid are filled.

With these pre-computed bins, performing a search for the nearest record simply involves querying the bin at the query point and all the bins in the 1-neighbourhood. This guarantees that distant and adjacent records are searched and it solves the problem encountered with the fixed grid method. It is important to note that this is only valid if we are querying inside the extents of the spatial index. If this is not the case, then we have to query the entire edge (all the bins along the side) of the spatial index for the nearest record.

The naïve approach to the described search method will search every bin in the neighborhood or every bin along the edge of the data extents. This, however, is not always necessary. If the distance of the nearest point in a bin to the query point is shorter than the distance to any of the bin edges, then we have found the nearest record and we do not have to search additional bins. This allows the algorithm to short circuit after searching through the first bin.

### D. Fixed Grid Spatial Index

By implementing the two abstract methods of the *AbstractGridSpatialIndex* to render the data, we would have a complete implementation for a fixed grid spatial index with no hierarchical sub-divisions. This is suitable for hand-tuned datasets or when a lightweight spatial index is required.

### E. Hierarchical Grid Spatial Index

The hierarchical grid spatial index is implemented by introducing a proxy collection (*SubGridDataCollection*) that implements *IGridSpatialIndexable,* and by overriding steps 6 and 10 of the abstract class' *RebuildIndex()* method. An internal list of all the sub grids is maintained and another grid stores the indices of these sub lists for each bin. When step 6 (*OnBeforeIndexRebuilt()*) is called, it merely recreates the sub grid lists. Step 10 (*OnAfterIndexRebuilt()*) does all of the actual work by going through all the unique integer lists and checking if their count exceeds *MaxBinRecords*. If this is the case then a clone of the current spatial index is made and it is passed a proxy to the integer list as its data source. This means that all the sub spatial indexes deal with a proxy to the original data source. This makes the implementation more efficient than making sub copies of the original data.

The hierarchical grid spatial index also overrides the *GetNearestRecord()* method in order to first check whether a sub grid needs to be queried. If this is the case then the query is passed down to the sub grid, otherwise the default implementation is used from the abstract base class.

A threshold parameter (*MaxBinRecords*) is used to decide when to sub divide a bin further with another Hierarchical Grid Spatial Index. Several schemes exist where the sub grids contain the same or varying amounts of sub divisions.

It is necessary to introduce a *SmallestBinDimension* parameter for this spatial index. This is because we need to limit the depth to which the spatial index will partition the search space. This is particularly important for the case when there are more than *MaxBinRecords* located at the exact same position. No matter how many times we sub divide the search space, we will never manage to partition the records any further. It is therefore important to have this threshold so that if either of the X or Y dimensions of the bins are smaller, then the partitioning stops.

## V. RESULTS AND ANALYSIS

In order to evaluate the search cost for the spatial index, an extent twice the size of the data extent is evaluated at a regular interval. The performance of the spatial index is evaluated for a varying number of grid divisions. This demonstrates what effect the regular grid has on spatial indexing (remember that this spatial index can be thought of as a bucket quad tree when the divisions are set to 2x2). The search cost in terms of number of records is evaluated at each point and the results are shown as an image. Two types of coloring schemes are used to look at the results. A relative color range is normalized to the minimum and maximum search costs for the image. This highlights areas of interest in the performance of the spatial index. An absolute range for the color is used so that the performance at different grid divisions can be compared. Examples of uniform and Gaussian data points are given in Fig. 7 and they show the corresponding search costs.

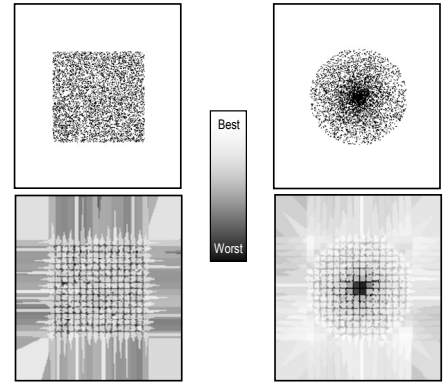

Fig. 7. Data points (top) and Search Costs (bottom). Left: 5000 Uniformly distributed points. Right: 5000 Gaussian points.

Fig. 8 shows the results for varying grid divisions for the grid spatial index. It is worthwhile looking at these results even though there is no hierarchical aspect to the algorithm, since the results can be thought of as a type of an impulse response for a particular layer.

When performing a nearest-point-search on a uniformly distributed set of data inside the data extents, our fixed grid spatial index should never exceed the maximum search cost given by (2):

$$c_{max} = 9 \times \frac{n}{(x \times y)} \qquad (2)$$

where n is the total number of records, x is the number of X divisions and y is the number of Y divisions. The reasoning behind this equation is that we need to search through the current bin plus 8 of its neighbors. For a uniform distribution, the average search cost is described by (1). The empirical results obtained so far (as seen in Fig. 8) show that this relationship is true. Equation (2) puts an upper bound on the search cost for the non-hierarchical grid spatial index. It also predicts that the maximum search cost decreases exponentially as the number of grid divisions increases. This clearly explains the decreasing trend in the graph of Fig. 8.

Fig. 9 shows the results for the hierarchical grid spatial index. The absolute value is scaled so that maximum color value corresponds to 1% of the total number of records. The *MaxBinRecords* threshold is set to 1 so that the grid is always sub-divided. We see that the performance of this spatial index is well under 1% so it is clear that it is effective at performing the spatial indexing tasks. The graph in Fig. 9 reaches the lower limit because of the *SmallestBinDimension* parameter.

In both hierarchical and non-hierarchical cases, the performance of spatial indexes with more than 2 divisions per dimension, the maximum search cost is always lower. This validates our previous expectation that better performance can be achieved by "looking" at the data with higher resolution images. There is an interesting memory trade off because having more division's means that the hierarchical tree will not be as deep as when there are only a few divisions per layer.

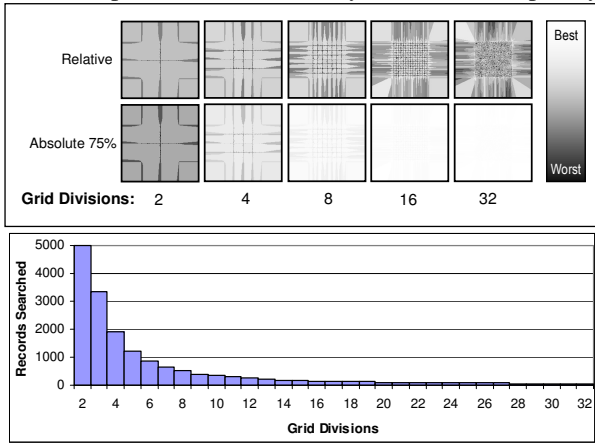

Fig. 8. Uniform Data Points (5000). No Hierarchical Divisions.

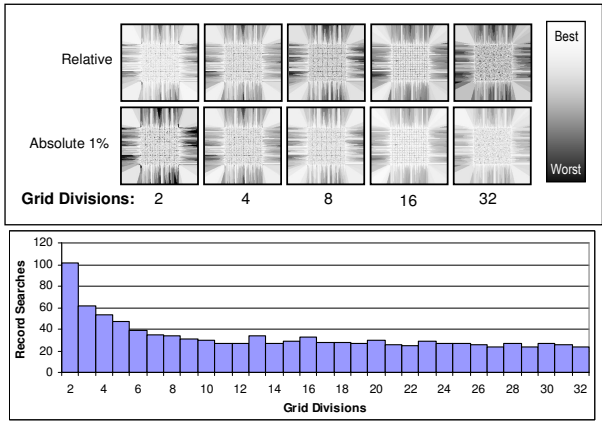

Fig. 9. Uniform Data Points (5000). Hierarchical Division, *MaxBinRecords* = 1.

## VI. FURTHER WORK

This paper has only analyzed the performance of this spatial indexing technique based on the number of records searched. Further work needs to be done to analyze the memory and time performance of the algorithm at varying grid divisions.

Since this spatial indexing method has roots in image processing, the algorithm is to be moved over to a hardware implementation where the indexing of the data is rendered by a hardware accelerated graphics card. This makes use of the card as a General Purpose Graphics Processing Unit (GPGPU) [11][12]. The algorithm will make use of the GPU to render the data to images that represent the grid in the spatial index described in this paper. The recent advances in vertex and pixel shaders on the GPU will make it feasible to implement a part of this spatial index on the hardware [13][14].

## VII. CONCLUSIONS

This paper describes the design and implementation of a hierarchical grid spatial index. It shows that treating the spatial indexing as an image processing operation makes for an elegant solution to the spatial indexing problem. The design of the classes has lead to an attractive solution where implementers of specific spatial indexes merely need to render the data onto a grid. The search costs for various grid divisions were analyzed and the results show that using more than 2 divisions per dimension (more than a quad tree) provides better search performance.